\documentclass[sigconf, nonacm]{acmart}
\usepackage{pvldb}
\usepackage{enumitem}
\usepackage{tabularx}
\makeatletter
\patchcmd{\@mkauthors@iii}{\par\medskip\leavevmode}{\par\leavevmode}{}{\errmessage{Author spacing patch failed}}
\makeatother
\setlist[itemize]{leftmargin=1.1em, itemsep=1pt, topsep=2pt, parsep=0pt}
\setlist[enumerate]{leftmargin=1.4em, itemsep=1pt, topsep=2pt, parsep=0pt}

\renewcommand\vldbdoi{XX.XX/XXX.XX}
\renewcommand\vldbpages{XXX-XXX}
\renewcommand\vldbavailabilityurl{https://github.com/hinsonver/CoRe-Benchmark}

\newcommand{\core}{\textsc{CoRe}}
\newcommand{\companion}{\href{https://github.com/hinsonver/CoRe-Benchmark/raw/refs/heads/main/paper/CoRe-Supplementary-Technical-Report.pdf}{supplementary report}}
\newcommand{\companionref}[1]{\href{https://github.com/hinsonver/CoRe-Benchmark/raw/refs/heads/main/paper/CoRe-Supplementary-Technical-Report.pdf\#nameddest=companion-#1}{Report~\S#1}}
\newcounter{finding}
\newcommand{\finding}[1]{\par\smallskip\noindent\refstepcounter{finding}%
  \fcolorbox{black!55}{black!4}{\parbox{\dimexpr\columnwidth-2\fboxsep-2\fboxrule\relax}{\small\textbf{Finding~\thefinding.} #1}}%
  \par\smallskip}

\extrafloats{40}
\begin{document}
\title[Cleaning or Recalling? Knowledge Dependence in LLM-based Data Cleaning]{Cleaning or Recalling? Knowledge Dependence in LLM-based Data Cleaning [Experiment, Analysis \& Benchmark]}

\author{Jingru Yang}
\affiliation[obeypunctuation=true]{\institution{Minzu University of China},~\country{China}}
\email{okiyang@muc.edu.cn}

\author{Honglu Zhang}
\affiliation[obeypunctuation=true]{\institution{Minzu University of China},~\country{China}}
\email{26302276@muc.edu.cn}

\author{Jiayi Mi}
\affiliation[obeypunctuation=true]{\institution{Minzu University of China},~\country{China}}
\email{25012086@muc.edu.cn}

\author{Ju Fan}
\affiliation[obeypunctuation=true]{\institution{Renmin University of China},~\country{China}}
\email{fanj@ruc.edu.cn}

\author{Zheng Chen}
\authornote{Zheng Chen is the corresponding author.}
\affiliation[obeypunctuation=true]{\institution{Minzu University of China},~\country{China}}
\email{202301050@muc.edu.cn}

\author{Xiaoyong Du}
\affiliation[obeypunctuation=true]{\institution{Renmin University of China},~\country{China}}
\email{duyong@ruc.edu.cn}

\AddToHook{shipout/lastpage}{\pdfinfo{/Author (Jingru Yang; Honglu Zhang; Jiayi Mi; Ju Fan; Zheng Chen; Xiaoyong Du)}}

\begin{abstract}
High scores on public data-cleaning benchmarks can reflect knowledge of familiar entities,
making it difficult to assess how well an LLM repairs unfamiliar data. We present \core{}
(\textbf{C}leaning \textbf{o}r \textbf{Re}calling), a protocol comparing public benchmarks with fresh real records and paired
name surrogates built from English words or syllables. The surrogates retain error positions
and non-identity values while transferring the benchmark's error mechanisms. Across 21
controlled tables and eight LLMs, repair performance varies substantially with entity names
on Hospital and, for commercial models, Movies. Renaming removes some information needed
for repair, such as a film's identifier mapping. It also affects repairs whose candidates
remain available: replacing entity names with word-based surrogates lowers
DeepSeek-V4-Pro's year accuracy from 91\% to 18\%.
On Hospital and Movies, commercial models score higher on fresh records than on word-based
surrogates. A screened continuation probe finds no positive evidence of benchmark-row memory
in the tested models.
Familiar facts can conflict with the task. Even with explicit table-evidence instructions,
commercial models rewrite assigned directors at rates of 0.000--0.404 on known films.
Reference scoring can also disagree with external records. On Rayyan,
an ISSN correction matching a title-matched PubMed record
receives no credit, whereas 94\% of TabClean's credited repairs reproduce the reference date
transformation. We provide controlled tables, scoring code and materials for offline replay.
\end{abstract}

\makeatletter
\let\core@savedtextbottom\@textbottom
\flushbottom
\AddToHookNext{shipout/after}{\global\let\@textbottom\core@savedtextbottom}
\makeatother
\maketitle

\vldbtopmatter

\section{Introduction}\label{sec:intro}
Data cleaning methods detect and repair erroneous values and are commonly evaluated on a small
set of public benchmark tables~\cite{rahm2000data,ilyas2019book,chu2016overview,abedjan2016detecting}.
Large language models (LLMs) have opened new approaches to data management~\cite{fernandez2023disrupt},
including few-shot error detection and imputation competitive with specialized
systems~\cite{narayan2022wrangle}. Recent cleaners use LLMs to edit values, derive rules or write
cleaning programs~\cite{ni2025zeroed,ni2024iterclean,zhang2025data,tabclean2026}. These methods,
however, can use information beyond the table being cleaned.

Public cleaning benchmarks contain familiar hospitals, films, breweries and articles.
Pretrained models can retrieve entity facts from their parameters~\cite{petroni2019lama,mallen2023trust},
while public tabular datasets can carry signals of prior exposure~\cite{bordt2024elephants,silvestri2025knowtable}.
A correct repair can therefore combine benchmark-row memory, knowledge of the entity
(\emph{entity knowledge}), knowledge of English word forms (\emph{spelling knowledge}), and
evidence within the table. Scores on familiar entities need not predict repair quality
on unfamiliar data.

\begin{figure}[tb]
  \centering
  \includegraphics[width=\columnwidth]{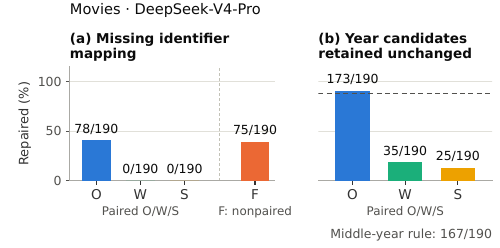}
  \caption{Recovery evidence in Movies. DeepSeek-V4-Pro repairs fewer identifiers when mappings
  disappear, but also fewer years when candidates remain. Labels give repaired/error cells;
  the middle-year rule matches 167/190 references. O/W/S are paired; F uses fresh records.}
  \label{fig:decomposition}
  \Description{Two Movies comparisons show fewer identifier repairs when mappings are removed and fewer year repairs despite retained year candidates.}
\end{figure}

\smallskip\noindent\textbf{Example 1.1.} Movies contains two distinct recovery problems.
A Rotten Tomatoes slug can stand in place of an IMDb identifier. For \emph{American Ultra},
recovering \texttt{tt3316948} from \texttt{american\_ultra} requires the film's mapping;
a fictitious slug does not determine its random identifier. Year ranges, however, retain the
same candidates after renaming. DeepSeek-V4-Pro's repairs decline in both cases
(Figure~\ref{fig:decomposition}), although the year candidates remain available.
\smallskip

Controlled tabular evaluations already test recognition and completion of dataset instances
(KnowTable~\cite{silvestri2025knowtable}) and adherence to counterfactual database values
(ContraTable~\cite{wang2026contratable}). Cleaning requires deciding which cells to change
and which source should supply their values. It also scores each edit against a reference,
which may itself contain a systematic conversion.

We introduce \core{} (\textbf{C}leaning \textbf{o}r \textbf{Re}calling) to compare repairs on dirty/reference pairs
under controlled entity replacements. Each benchmark is compared with \emph{fresh} real records,
a \emph{word-based surrogate} with fictitious English-word names, and a \emph{syllable-based surrogate}
with meaningless names. The original and its surrogates share rows and error positions, while
fresh records carry the same kinds of errors. Paired substitutions test entity familiarity
and word form. Fresh records test performance beyond the benchmark, while a separate
continuation probe tests for benchmark-row memory.

We transfer the benchmarks' error mechanisms to new records and renamed identity values,
retaining error positions and non-identity values in the paired surrogates. This is particularly
important for Movies, where errors
often reflect differences between two legitimate sources. Fictitious names must also hide the
entity without themselves looking erroneous. We build 21 controlled tables: nine each for
Hospital and Movies across three seeds, and three for Beers. Their surrogates contain about 41,000 fictitious
values, assessed by 506 acceptance checks. We evaluate four commercial and four open-weight
LLMs, alongside Baran~\cite{mahdavi2020baran} and the program-generating cleaner
TabClean~\cite{tabclean2026}, using about 23,000 prompts. Our main findings are as follows.

Movie identifiers lose their mappings after renaming, but year ranges retain the same candidates.
On Movies, DeepSeek-V4-Pro's year accuracy falls from 91\% on real films to 18\% on word names
and 13\% on syllable names, even though the candidates are unchanged
(Figure~\ref{fig:decomposition}). Across the four commercial models, Movies' identity-column
F\,$-$\,W gaps are $+$0.17 to $+$0.42 at construction seed~0 (Table~\ref{tab:main}).
On Hospital, three commercial models lose 0.59 to 0.72 repair F1 after syllable replacement
on \emph{identity columns}, which contain entity names and locations.

The screened continuation probe finds no positive evidence of benchmark-row memory in its tested
models (Section~\ref{sec:memorization}). Hospital and Movies retain positive condition contrasts
across construction seeds and, for DeepSeek-V4-Pro, with labeled examples and error positions.
Hospital's O\,$-$\,S gap grows across the three Qwen2.5 model sizes. On Hospital and Beers,
whole-table Baran and TabClean retain high scores under renaming, with repeated records and
formatting patterns available for repair (Sections~\ref{sec:knowledge}--\ref{sec:generalize}).
The \companion{}\footnote{\url{https://github.com/hinsonver/CoRe-Benchmark/blob/main/paper/CoRe-Supplementary-Technical-Report.pdf}.} gives the overview with intervals.

Yet a familiar fact can be the wrong value to preserve. Under explicit table-evidence
instructions, commercial models rewrite 0.000--0.404 of assigned directors of known films
toward real-world credits, compared with 68\%--100\% under the standard prompt. Qwen2.5-14B
still rewrites more than 60\% under the explicit instruction. Hospital and brewery city
rewrites are less frequent (Section~\ref{sec:conflict}).

Reference errors can also change which edits receive credit. Rayyan contains systematic date and page
conversions (Section~\ref{sec:quality}). A repair that matches a title-linked PubMed ISSN
receives no credit because the reference contains a spreadsheet conversion. On TabClean's
input, 434 of 461 credited repairs (94\%), after excluding development rows, reproduce the
reference date transformation. Stratified PubMed field checks examine these references
separately, with field-specific criteria and measured title-match coverage.

\noindent We make the following contributions:
\begin{itemize}
  \item We propose \core{}, a controlled evaluation protocol that transfers benchmark errors
  to paired name surrogates and relates repair scores to the recovery evidence retained in
  the input (Section~\ref{sec:protocol}).
  \item We construct controlled tables for three benchmarks and provide construction and
  evaluation code, numerical witnesses and replay materials
  (Sections~\ref{sec:construction} and~\ref{sec:artifact}).
  \item We characterize both knowledge-dependent repair and directed rewriting of assigned
  values across eight LLMs, with separate continuation and information-condition checks
  (Sections~\ref{sec:memorization}--\ref{sec:conflict}).
  \item We identify systematic reference transformations in Rayyan and trace how they receive
  repair credit (Section~\ref{sec:quality}).
\end{itemize}

Section~\ref{sec:related} provides background and related work.
Sections~\ref{sec:protocol}--\ref{sec:setup} describe the protocol, constructions and evaluation
settings. Sections~\ref{sec:memorization}--\ref{sec:quality} present the analyses, followed by
recommendations and limitations in Section~\ref{sec:discussion}.

\section{Background and Related Work}\label{sec:related}

\subsection{Cleaning Benchmarks}\label{sec:benchmarks}

Cleaning systems draw on several kinds of evidence. Rules repair constraint violations or
apply fixing patterns~\cite{dallachiesa2013nadeef,chu2013holistic,wang2014fixing}, while KATARA
uses knowledge bases and crowdsourcing~\cite{chu2015katara}. Learned detectors use a small
set of labels~\cite{mahdavi2019raha,heidari2019holodetect,neutatz2019ed2}; Baran learns context
models for repair~\cite{mahdavi2020baran}, and HoloClean combines signals through probabilistic
inference~\cite{rekatsinas2017holoclean}. Experimental studies compare detection and repair
tools~\cite{abedjan2016detecting,ni2024repair} and their effects on machine-learning
models~\cite{li2021cleanml,abdelaal2023rein}. ActiveClean integrates progressive cleaning with
statistical model training~\cite{krishnan2016activeclean}. BART controls detectability and
repairability during error generation~\cite{arocena2015bart}.

We select public dirty/reference pairs from Raha and Baran for their complementary error
profiles (Table~\ref{tab:datasets}). Hospital's damaged names contrast with Movies' source-specific
mappings and retained candidates. Beers adds a formatting contrast: its errors retain the
information needed for repair. Rayyan's spreadsheet
transformations instead expose what reference matching rewards. For cleaning we use 1,000-row
tables. Movies and Beers retain all source-replaced and city-error rows, respectively, plus
sampled benchmark rows. Flights and Tax serve only the continuation probe.

\begin{table*}[t]
  \caption{Benchmarks distributed with Raha and Baran. Parentheses give identity-column error
  cells. CMS: US Centers for Medicare \& Medicaid Services.}
  \label{tab:datasets}
  \small
  \setlength{\tabcolsep}{4pt}
  \begin{tabular*}{\textwidth}{@{\extracolsep{\fill}}lllrll@{}}
    \toprule
    Dataset & Entities & Source & Error cells & Kind of error & Use in this paper \\
    \midrule
    Hospital & US hospitals & CMS registry & 509 (219) & one character replaced by \texttt{x} & four conditions \\
    Movies & films & IMDb and Rotten Tomatoes & 1,965 (379) & the other source's value; formatting rules & four conditions \\
    Beers & craft beers and breweries & CraftCans.com & 565 (127) & state glued to city; alcohol \% suffix & formatting control \\
    Rayyan & biomedical articles & PubMed & 948 & spreadsheet conversions (Section~\ref{sec:quality}) & quality analysis \\
    Flights & flight times & 38 web sources & \multicolumn{2}{l}{not used for cleaning} & memorization probe \\
    Tax & personal tax records & --- & \multicolumn{2}{l}{not used for cleaning} & memorization probe \\
    \bottomrule
  \end{tabular*}
\end{table*}

\emph{Identity columns} name or locate an entity: provider number, name, address, city, ZIP code,
county and phone in Hospital; identifier, title, director, writers and actors in Movies; and beer,
brewery and city in Beers. We distinguish their recovery from repairs of other values, such as
film years and formatting errors.

\subsection{LLM-Based Cleaners}\label{sec:cleaners}

Few-shot prompting can match specialized systems on studied detection, imputation and matching
tasks~\cite{narayan2022wrangle}. Subsequent work includes table-tuned and preprocessing
models~\cite{li2024tablegpt,zhang2024jellyfish} and entity
matchers~\cite{tu2023unicorn,peeters2025entity}. LLM cleaners label errors
(ZeroED~\cite{ni2025zeroed}), iterate on repairs (IterClean~\cite{ni2024iterclean}), derive
ontological or semantic rules (LLMClean~\cite{10.1007/978-3-031-70421-5_7},
Cocoon~\cite{zhang2025data}), standardize data as agents (CleanAgent~\cite{qi2024cleanagent}),
compile curation pipelines (SEED~\cite{chen2024seeddomainspecificdatacuration}), or generate
cleaning programs (TabClean~\cite{tabclean2026}). We study direct value editing and generated
programs under controlled entity replacements.

\begingroup
\makeatletter
\patchcmd{\subsection}{-.75\baselineskip}{-2pt}{}{\errmessage{Local subsection spacing patch failed}}
\makeatother
\subsection{Knowledge and Benchmark Evaluation}\label{sec:relatedwork}

\textbf{LLMs on enterprise data.} Enterprise annotation and integration studies examine domain
context, schemas, table width and sparsity~\cite{kayali_cidr25,demiralp2024enterprise}.
Bodensohn et al. also observe a bias toward familiar SQL despite explicit SIGNAL
documentation~\cite{bodensohn2025enterprise}.
\par
\endgroup

\textbf{Memorization and contamination.} LLMs can memorize training text, with stronger effects
as models grow or data repeat~\cite{274574,carlini2023quantifying}. Detection methods include
benchmark completion~\cite{ICLR2024_bc39a59c}, likelihoods of canonical and shuffled orderings
under exchangeability~\cite{oren2024proving}, and least-likely token
probabilities~\cite{shi2024detecting}. Benchmark-specific records and surveys document these
issues~\cite{sainz2023trouble,xu2024contamination}, while mitigation restricts exposure or
refreshes questions~\cite{jacovi2023stop,white2025livebench}.
For tables, Bordt et al.~\cite{bordt2024elephants} find strong memorization of Iris and Wine;
Adult's signals concentrate in headers and initial rows. KnowTable~\cite{silvestri2025knowtable}
compares completion and existence queries over real, column-resampled, swapped and obfuscated
variants, with non-neural baselines and statistical tests. Our comparison instead follows
dirty/reference cell pairs: we transfer their error mechanisms to renamed entities and relate
repair scores to the candidates and patterns left in the input. The continuation probe separately
uses structural baselines to distinguish predictable cells from evidence of row memory.

\textbf{Knowledge and counterfactual evaluation.} Models store facts about entities, especially
popular ones~\cite{petroni2019lama,mallen2023trust,3618408.3619049}. Entity substitutions in QA
reveal conflicts between parametric and contextual answers~\cite{longpre2021conflicts,xie2024adaptive,xu2024conflictsurvey},
and counterfactual or low-probability variants expose dependence on familiar task
instances~\cite{wu2024reasoning,mccoy2024embers,ICLR2025_ec2e7a89}. DisentQA separates contextual
and parametric answers~\cite{neeman2023disentqa}. In data management,
ContraTable~\cite{wang2026contratable} pairs original and counterfactual databases to compare
answer faithfulness across lookup, joins and temporal reasoning. A cleaner must both change
errors and preserve assigned values. Our conflict test measures directed rewrites of
reference-consistent cells toward familiar facts, alongside arbitrary edits
(Section~\ref{sec:conflictmetric}). Fictional ontologies and pseudowords also limit factual shortcuts
in reasoning tasks~\cite{saparov2023language,derelab2026}. We contrast word and syllable names
for cleaning.

\textbf{Benchmark quality.} Label errors can change model comparisons, particularly on
mislabeled subsets and at higher error prevalence~\cite{northcutt2021pervasive}. Spreadsheet
conversion can also silently alter values~\cite{ziemann2016gene}. Rayyan lets us trace credited
repairs back to reference transformations and examine what the score rewards.

\section{The \core{} Protocol}\label{sec:protocol}

\begin{figure*}[t]
  \centering
  \includegraphics[width=\textwidth]{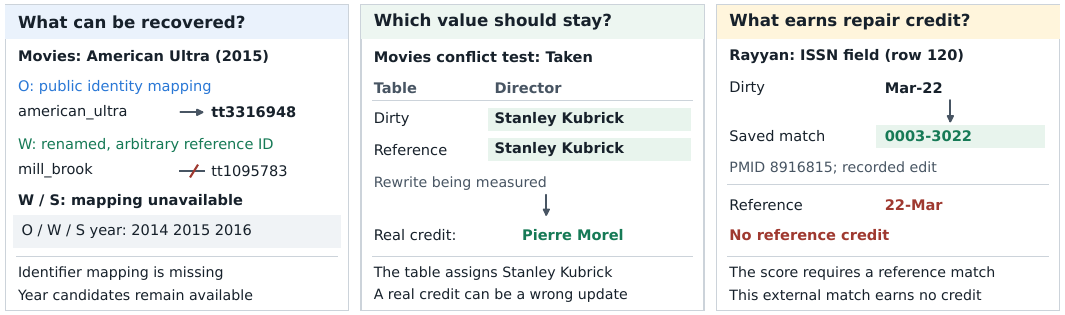}
  \caption{Three questions for interpreting repair scores: what evidence remains for recovery,
  which assigned values should be preserved, and which edits receive reference credit.
  The panels illustrate separate tests on their respective benchmark inputs.}
  \label{fig:overview}
  \Description{Three separate examples illustrate recovery evidence, preservation of an assigned director, and an externally matched ISSN correction rejected by reference scoring.}
\end{figure*}

\core{} evaluates repair under four input conditions. Within a setting, the prompt,
information access and scoring are fixed. Paired surrogates replace identity values in the
same dirty/reference rows; fresh records provide a separate comparison on other records.
Figure~\ref{fig:overview} also shows two checks beyond recovery: whether a cleaner preserves an
assigned director, and whether a repair agrees with the scoring reference. These use the
conflict and reference inputs in Sections~\ref{sec:conflict} and~\ref{sec:quality}.

\subsection{Four Conditions and Their Comparisons}\label{sec:conditions}

\begin{figure*}[t]
  \centering
  \includegraphics[width=\textwidth]{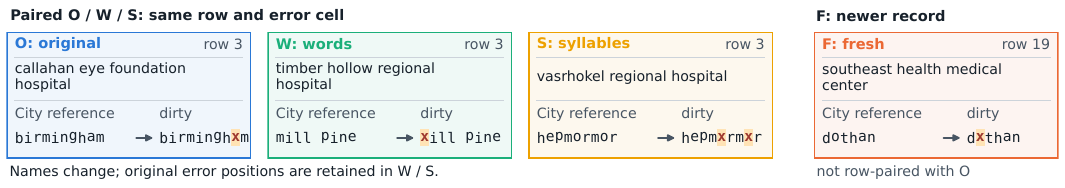}
  \caption{Hospital construction example. O/W/S share a row and city error cell while
  changing identity values. F uses a newer registry row with a newly drawn error location.
  The \companion{} gives the extended identity-field example.}
  \label{fig:example}
  \Description{Hospital rows compare original, word and syllable names with a shared city error; a fresh registry row uses a newly drawn error location.}
\end{figure*}

Besides the original (\textbf{O}), \core{} builds three tables with the same schema and size:
\begin{itemize}
  \item \textbf{Fresh (F)} uses real records outside the benchmark: newer registry records for
  Hospital, or other films and beers. Their entities may remain familiar to a model.
  \item \textbf{Word-based surrogate (W)} replaces identity values one to one with fictitious names
  made of English words and random numeric identifiers, retaining non-identity values and
  error positions.
  \item \textbf{Syllable-based surrogate (S)} uses syllable names in the same construction, retaining
  generic terms such as \emph{hospital} and \emph{street}.
\end{itemize}
Figure~\ref{fig:example} illustrates naming and error transfer on Hospital rows.

\begin{table*}[t]
  \caption{Condition contrasts and their interpretation. O/W/S pair rows, error cells, chunks
  and non-identity values; F shares schema, size and error recipes or rules.}
  \label{tab:conditions}
  \small
  \begin{tabularx}{\textwidth}{@{}l >{\raggedright\arraybackslash}X@{}}
    \toprule
    Contrast & Interpretation and comparison assumption \\
    \midrule
    O$-$W & Paired entity-renaming contrast; replacement changes familiarity, word forms and factual associations between fields. \\
    W$-$S & Word-form contrast; interpretation depends on the difficulty of the two naming styles and any regularities they introduce. \\
    O$-$S & Paired O--S contrast, combining changes in entity familiarity and word form. \\
    F$-$W & Fresh records versus word-based surrogates; entity familiarity changes with record difficulty and, where recipes cannot transfer exactly, error composition. \\
    O$-$F & Original--fresh comparison; benchmark exposure and table difficulty both change. Memorization is tested separately. \\
    \bottomrule
  \end{tabularx}
\end{table*}

Let $s_O,s_F,s_W,s_S$ denote a method's repair F1 in a fixed column group.
Table~\ref{tab:conditions} defines three paired contrasts: O\,$-$\,W, W\,$-$\,S and O\,$-$\,S.
They measure score changes under entity and word-form replacement; the first two sum to the third.
We interpret these differences alongside the error profiles and information-access controls,
since renaming changes candidate values and
construction regularities as well as familiarity. F\,$-$\,W compares fresh real records with
word-based surrogates and is reported separately. O\,$-$\,F also changes record difficulty;
the continuation probe examines benchmark-row memory separately.

\subsection{What Makes an Error Recoverable?}\label{sec:recoverability}

Recovery depends on the target convention and the information shown to the cleaner
(Table~\ref{tab:recoverability}). The reference specifies the target convention, which is not
explicitly stated in the zero-shot prompt. A \emph{locally reversible} error admits a conversion to the
specified convention (R1). A \emph{constrained} error retains candidates or regularities that
narrow the choices (R2), without necessarily selecting a unique source value.
An \emph{underdetermined} error lacks a required mapping or replacement value (R3).
We describe these properties for each error type and information setting, not from a model's
success rate. These properties can overlap.

Where rules or candidates remain available, paired contrasts test their use after renaming.
When the required mapping is absent, lower scores instead expose the recovery evidence removed
by construction. A repeated record or labeled example can supply a previously missing value
(Section~\ref{sec:labels}); a candidate list need not resolve which source takes priority.
\companionref{4} gives the criteria and error-type coverage. The S score includes repairs
using local rules and construction regularities.

\begin{table}[t]
  \caption{Recovery evidence relative to the shown input and target convention, as defined in
  Section~\ref{sec:recoverability}.}
  \label{tab:recoverability}
  \small
  \begin{tabularx}{\columnwidth}{@{}>{\raggedright\arraybackslash}p{0.25\columnwidth} >{\raggedright\arraybackslash}X@{}}
    \toprule
    Case & Examples in the benchmark conditions \\
    \midrule
    R1: locally reversible & Beers city/state split and percentage suffix; Movies thousands separators and integer ratings with an added \texttt{.0}. \\
    R2: constrained & Movies year ranges; Hospital name typos with a matching candidate; syllable names with a restricted construction vocabulary. \\
    R3: underdetermined & Random fictitious identifiers; omitted film credits; source-specific ratings and durations not recoverable from the other source. \\
    \bottomrule
  \end{tabularx}
\end{table}

\subsection{Chunking, Prompting and Scoring}\label{sec:scoring}

The main input contains 20 rows from different entities, preventing copying from another record
of the same hospital or brewery while retaining cross-entity patterns. The same entity grouping
gives O/W/S identical chunks, 50 per table. A zero-shot prompt requests JSON corrections with
row, column and value fields, leaving correct cells unchanged. Other information
settings are defined in Section~\ref{sec:setup}.

Let $E$ be the cells whose dirty and reference strings differ, $P$ the cells effectively edited
after no-ops are removed, and $R \subseteq P \cap E$ those set to the reference value. Then
\begin{equation*}
  \textit{prec} = \frac{|R|}{|P|}, \qquad \textit{rec} = \frac{|R|}{|E|}, \qquad
  F1 = \frac{2 \cdot \textit{prec} \cdot \textit{rec}}{\textit{prec} + \textit{rec}},
\end{equation*}
and detection F1 uses $P \cap E$ in place of $R$. Edit and repaired-value comparisons ignore
case and normalize whitespace, while $E$ uses original strings. Thus Movies' five case-only
title errors remain in $E$ but cannot earn repair credit. Identity columns are reported
separately from the others. Normalization and reply handling follow \companionref{2};
the exact prompt is in \companionref{7}.

\section{Building the Controlled Tables}\label{sec:construction}

The controlled tables retain the benchmark's error mechanisms while varying entities and word
forms. Error generators can vary detectability and repairability~\cite{arocena2015bart}; here,
we start from the observed error profile and transfer it to renamed values or fresh records.
Surrogates preserve non-identity cells and error positions, whereas fresh error counts depend
on whether recipes and rules apply to the new records. Naming rules hide identity cues while
retaining expected word forms. Table~\ref{tab:construction}
summarizes the constructions, with full recipes and checks in \companionref{1}.

\begin{table*}[t]
  \caption{Controlled-table sources and scale. Parentheses give error counts in identity columns;
  values renamed are distinct one-to-one replacements. Seeds cover F, W and S.}
  \label{tab:construction}
  \small
  \begin{tabular*}{\textwidth}{@{\extracolsep{\fill}}l lr llr rc@{}}
    \toprule
    & \multicolumn{2}{c}{Original (O)} & \multicolumn{3}{c}{Fresh (F)} & \multicolumn{1}{c}{W and S} & \\
    \cmidrule(lr){2-3}\cmidrule(lr){4-6}\cmidrule(lr){7-7}
    Dataset & Entities & Errors & Source & Entities & Errors & Values renamed & Seeds \\
    \midrule
    Hospital & 45 hospitals & 509 (219) & CMS registry, July 2026 release & 34 hospitals (33 also in O) & 509 (219) & 296 & 3 \\
    Movies & 1,000 films & 1,965 (379) & IMDb datasets, Sept.\ 2026 & 1,000 films & 2,195 (371) & 6,018 & 3 \\
    Beers & 421 breweries & 565 (127) & Wikidata, Sept.\ 2026 & 94 breweries & 565 (127) & 1,712 & 1 \\
    \bottomrule
  \end{tabular*}
\end{table*}

\subsection{Hospital}\label{sec:build-hospital}

Hospital's characteristic corruption replaces a character by \texttt{x}: 508 of 509 error cells
follow this form, usually replacing every occurrence of that character. F uses the July 2026
CMS registry in the benchmark schema. Its 34 hospitals include 33 from O, but all quality
measures have changed, giving new rows about largely the same entities. We retain the original
per-column error counts and position-weighted substitution rule, with newly drawn error locations.

Surrogates rename each distinct identity value consistently and transfer the original errors
by position (Figure~\ref{fig:example}). S uses an inventory of 24 syllables; W uses English-word
combinations. Generic suffixes preserve column conventions, and numeric identifiers become
random digits. Column-wise replacements preserve repeated-value relationships, but random
ZIP codes and phones are not constrained by the retained State values. Thus Hospital also
changes factual associations between fields; \companionref{1} details these relationships
and the archived-edit checks. The restricted syllable inventory supplies a regularity examined
in Section~\ref{sec:commercial}.

\subsection{Movies}\label{sec:build-movies}

Movies joins IMDb and Rotten Tomatoes records. Its 190 source-replaced rows, together with two
formatting rules, produce the errors in Table~\ref{tab:movieserrors}. A film can have a Rotten
Tomatoes slug instead of an IMDb identifier, a year range instead of a year, or credits that
add or omit people. For example, \emph{American Ultra} has \texttt{american\_ultra} in place of
\texttt{tt3316948}. Such differences require recovering the benchmark's preferred source value,
rather than merely fixing spelling.

\begin{table*}[t]
  \caption{Movies' 1,965 errors and their transfer mechanisms. Source-style errors occupy
  190 rows, with formatting rules also applied table-wide. Copied errors transfer cell recipes;
  drawn errors follow measured rates. Slugs are the film identifiers used in Rotten Tomatoes URLs.}
  \label{tab:movieserrors}
  \small
  \begin{tabular}{@{}lr p{6.4cm} p{3.75cm} p{3.75cm}@{}}
    \toprule
    Column & Errors & In the original & Fresh table & Surrogates \\
    \midrule
    identifier & 190 & Rotten Tomatoes slug (135 plain, 28 numbered, 27 with the year) & slug of the fresh title, forms at the original's rates & slug of the dirty fictitious title, same form \\
    year & 190 & range ``Y$-$1 Y Y$+$1'' & same range & the original's cell \\
    title & 17 & 12 edit recipes (capitalization, \& for \emph{and}, accent, punctuation, subtitle, article) & the same edits on a fresh title & the same edits on the fictitious title \\
    director & 43 & 115 names added, 1 dropped, 17 respelled & copied name by name & copied name by name \\
    writers & 129 & 123 names added, 51 dropped, 20 respelled; 5 cells emptied & copied name by name & copied name by name \\
    duration & 187 & Rotten Tomatoes format (105) or empty (80) & drawn at the original's rates & the original's cell \\
    rating & 261 & score pair (165) or empty (22); ``.0'' added (74) & drawn at the original's rates & the original's cell \\
    rating count & 768 & thousands separators on every count $\geq$ 1,000 & the same rule & the original's cell \\
    genre & 180 & Rotten Tomatoes genre vocabulary & copied slot by slot & the original's cell \\
    \bottomrule
  \end{tabular}
\end{table*}

F matches each original film to another feature film in the September 2026 IMDb snapshot,
using the same release year (or within two years) and the nearest logarithmic vote count among
eligible films with at least 1,000 votes. The corresponding 190 rows receive source-style errors.
People and title errors are transferred as recipes over names and edits; genre errors retain
the other source's vocabulary. A recipe that cannot apply to its assigned cell is moved to
another source-style row able to take it.

Some recipes remain inapplicable: four director and four writer errors stay clean, leaving 371
identity errors in F versus 379 in O/W/S. Formatting rules also change counts: the separator
rule affects all 1,000 fresh rating counts, compared with 768 in O. F therefore has 2,195 errors
at seed~0 versus 1,965 in O/W/S. Table~\ref{tab:conditions} specifies the nonpaired F comparison.

Surrogates rename the 6,018 distinct identity values one to one. Titles require special care:
retaining a real title's function words, numbers and punctuation can reveal the film. We
generate title shapes independently at the original's aggregate rates, exclude real title
words from W, and require generated titles to support their transferred edit recipes. Fictitious
IMDb identifiers and slugs are then formed consistently with those titles.

Recovery evidence varies even within a column. Of 105 duration errors expressed in hours and
minutes, unit conversion agrees with the reference in 50 and disagrees in 55. A locally available
conversion can therefore differ from the preferred source value. Year ranges instead retain
useful candidates, whose use we examine in Section~\ref{sec:knowledge}.

\subsection{Beers}\label{sec:build-beers}

All Beers errors retain the information needed for formatting repair. They append a percentage
sign to a fractional alcohol value, or glue a state to a city and empty the state field. Removing
the suffix or splitting the location restores the benchmark's convention. F uses 1,000 beers
from 94 Wikidata breweries~\cite{vrandecic2014wikidata}, excluding original beers and transferring
the original error forms and counts.

Surrogates preserve generic terms and consistent relationships between brewery and city names.
A one-word city stays one word, avoiding the appearance of a glued city/state value.

\subsection{Acceptance Checks and Construction Seeds}\label{sec:acceptance}\label{sec:naming}

Acceptance checks cover structure, error counts and strength, naming leakage, fresh-source
requirements and deterministic rebuilding. The tables pass 505 of 506 checks. Hospital W at
seed~0 has fewer errors replacing three or more characters: 6.4\%, versus 16.9\% in O.
The 10.5-point difference exceeds the eight-point tolerance by 2.5 points; error positions
are unchanged. \companionref{1} gives the thresholds and \companionref{7} the distributions.

Naming and leakage account for 263 checks: one-to-one replacements, exclusion of real values,
and title-shape comparisons against chance-level matching. Deterministic rebuilding checks
that the same rules yield identical tables.

Hospital and Movies have three construction seeds: surrogates redraw names and applicable
errors; F redraws errors on fixed clean records. Seed ranges capture construction variation,
whereas cluster intervals describe one fixed construction. Main tables report seed~0.
Hospital W at seeds 1 and 2 passes the strength check, with positive F\,$-$\,W and
W\,$-$\,S contrasts for all eight cleaners. Table~\ref{tab:seeds} gives construction ranges;
\companionref{7} gives the passing-seed contrasts.

\section{Experimental Setup}\label{sec:setup}

\textbf{Models.} Commercial requests disable thinking and set temperature~0 for
DeepSeek-V4-Pro~\cite{deepseekai2026deepseekv4} and Qwen3.8-Max~\cite{qwen38}.
GLM-5.3~\cite{zai2026glm53} uses low reasoning effort and temperature~0;
GPT-6.1-Sol~\cite{openai2026gpt61sol} uses low reasoning effort with sampling parameters
left at API defaults. Settings stay fixed across conditions within each model.
Locally, vLLM~\cite{kwon2023vllm} greedily decodes
Qwen2.5 Instruct at 7B, 14B and 32B (AWQ)~\cite{qwen2025qwen25}, and
Llama-3.1-8B-Instruct~\cite{dubey2024llama3}. OLMo-2-7B instruction/base
variants~\cite{olmo2024olmo2} support auxiliary continuation with a public pretraining corpus.
Table labels shorten model names; Qwen-7B/14B/32B denote Qwen2.5.
Eight of GLM's 500 Movies replies remain incomplete with no parsed corrections.
\companionref{2} gives budgets and retries; \companionref{7} covers usage and sensitivity.

\textbf{Methods and information.} The \emph{aligned} setting adds 20 labeled dirty/clean tuples
and error positions to the main chunks. Baran~\cite{mahdavi2020baran} uses the same aligned
inputs with three seeds, plus whole tables with perfect detection or
Raha~\cite{mahdavi2019raha}. We disable Baran's Wikipedia-revision value models and Raha's
knowledge-base detector. TabClean runs DeepSeek-V4-Pro on whole tables with repeated records;
scoring excludes 150 labeled development rows.

\textbf{Scale.} The 1,000-row inputs retain Movies' source-style and Beers' city-error rows
(Section~\ref{sec:benchmarks}; \companionref{1}). Fixed paired O/W/S chunks and Hospital/Movies
construction-seed repetitions support recovery-mechanism comparisons
(Section~\ref{sec:acceptance}). \companionref{2} details about 23,000 cleaning,
aligned, conflict and continuation prompts; Rayyan input versions follow Section~\ref{sec:quality}.

\textbf{Statistics.} Repair intervals use 50,000 bootstrap reweightings~\cite{efron1993bootstrap,owen2012arrays}.
Hospital and Beers multiply independent entity and reply weights; Movies reweights replies.
O/W/S share paired weights, while F uses independent reply weights. Weighted counts are
pooled before computing F1. Two-sided tests use centered bootstrap differences, with
Holm correction~\cite{holm1979simple} within each result table (40 contrasts in
Table~\ref{tab:main}). Chunk wins/losses are descriptive where entities recur;
Movies sign tests form a separate eight-model family.

Conflict intervals resample whole replies 50,000 times, retaining equal entity weights
(Section~\ref{sec:conflictmetric}); all-or-none outcomes are reported as counts and rates.
Rayyan uses 5,000 within-stratum resamples of saved binary checks with fixed type weights,
conditional on title matches and field criteria (Section~\ref{sec:rayyan-audit}).
Intervals use the 2.5th and 97.5th percentiles; \companionref{7} details the clustering.

\section{Testing Benchmark Memorization}\label{sec:memorization}

The continuation probe covers DeepSeek-V4-Pro, the four open-weight cleaners and both OLMo-2
variants. It asks for the next row after a header and ten consecutive rows, and tests whether
dirty context elicits a clean reference value. \companionref{2} details trial counts and context settings.

\textbf{Screening predictable continuations.} Repeated values and record order make many
continuations predictable. A cell is \emph{hard} if its value is absent from the context column
and missed by a full-table screen: column modes, frequent successors, low-cardinality functional dependencies
and previous-value $\pm1$. Counts subtract the target vote; dependency eligibility stays fixed.
The screen uses records beyond the prompt (\companionref{2}). On Hospital, screening lowers
DeepSeek's context-absent accuracy from 47\% to 2.1\% (3/144).

The highest hard-cell accuracy is 32/375 (0.085; Table~\ref{tab:p1}). On Beers, many replies name
other real beers of the contextual brewery, not the next beer in the file.
OLMo-2 base also has low accuracy (at most 0.011); an
infini-gram~\cite{liu2024infinigram} search of its public corpus finds only Hospital fragments
in a GitHub issue and tool documentation, and none of the other tables.

\textbf{Clean-value leakage.} For clean references absent from the corresponding context column,
the hard-cell screen retains 601 erroneous cell-trials for DeepSeek, which supplies the clean
reference in three; the four open instruction cleaners do so in 1 of 2,008.
This additional screen uses the full clean table and clean predecessor---oracle
information beyond the dirty prompt.

\textbf{Original versus fresh.} All four commercial cleaners score lower on O than F on
Hospital ($-$0.03 to $-$0.07) and Movies ($-$0.09 to $-$0.25; Table~\ref{tab:main}). Some open
models have the opposite sign, including Llama's $+$0.10 on Hospital, without positive evidence
from the continuation probe. As defined in Table~\ref{tab:conditions}, O\,$-$\,F also includes difficulty changes
in addition to exposure. Section~\ref{sec:knowledge} compares repair on these real records with
the paired surrogates.

\finding{The screened continuation results provide no positive evidence attributable to
benchmark-row memory in the tested settings.}\label{find:mem}

\begin{table}[t]
  \caption{Highest hard-cell continuation accuracy over datasets for each context.
  Columns distinguish clean/dirty context, dataset naming and error-row selection.
  Leakage counts clean reference values written from dirty context.}
  \label{tab:p1}
  \small
  \setlength{\tabcolsep}{2.5pt}
  \begin{tabular}{@{}lcccccr@{}}
    \toprule
    & \multicolumn{2}{c}{Anonymous} & \multicolumn{2}{c}{Named} & Error & Clean value \\
    \cmidrule(lr){2-3}\cmidrule(lr){4-5}
    Model & clean & dirty & clean & dirty & rows & leaked \\
    \midrule
    DeepSeek-V4-Pro & 0.080 & 0.070 & 0.085 & 0.065 & --- & 3 / 601 \\
    OLMo-2-7B-Instruct & 0.027 & 0.019 & --- & --- & 0.030 & 0 / 187 \\
    OLMo-2-7B (base) & 0.011 & 0.011 & --- & --- & --- & 1 / 92 \\
    Qwen2.5-7B & 0.040 & 0.028 & 0.040 & 0.046 & 0.029 & 0 / 449 \\
    Llama-3.1-8B & 0.042 & 0.019 & 0.014 & 0.009 & 0.016 & 1 / 490 \\
    Qwen2.5-14B & 0.027 & 0.061 & 0.041 & 0.049 & 0.030 & 0 / 532 \\
    Qwen2.5-32B & 0.005 & 0.037 & 0.013 & 0.028 & 0.029 & 0 / 537 \\
    \bottomrule
  \end{tabular}
\end{table}

\section{Entity Familiarity and Repair Performance}\label{sec:knowledge}

Table~\ref{tab:main} reports the eight-model condition scores and contrasts. The error profiles
in Section~\ref{sec:recoverability} distinguish missing entity mappings from candidates and rules
retained after renaming.

\begin{table*}[t]
  \caption{Identity-column repair F1 at seed 0. F$-$W uses nonpaired fresh records; O/W/S contrasts are paired. $^\dagger$: Holm-corrected $p<0.05$ over 40 contrasts, using 50,000 entity/reply bootstrap reweightings (Section~\ref{sec:setup}).}
  \label{tab:main}
  \small
  \begin{tabular*}{\textwidth}{@{\extracolsep{\fill}}l cccc ccc cccc cc@{}}
    \toprule
    & \multicolumn{7}{c}{Hospital} & \multicolumn{6}{c}{Movies} \\
    \cmidrule(lr){2-8}\cmidrule(l){9-14}
    & \multicolumn{4}{c}{Scores} & Nonpaired & \multicolumn{2}{c}{Paired} & \multicolumn{4}{c}{Scores} & Nonpaired & Paired \\
    \cmidrule(lr){2-5}\cmidrule(lr){6-6}\cmidrule(lr){7-8}\cmidrule(lr){9-12}\cmidrule(lr){13-13}\cmidrule(l){14-14}
    Model & O & F & W & S & F$-$W & W$-$S & O$-$S & O & F & W & S & F$-$W & O$-$W \\
    \midrule
    DeepSeek-V4-Pro & 0.88 & 0.95 & 0.50 & 0.16 & +0.45$^\dagger$ & +0.34$^\dagger$ & +0.72$^\dagger$ & 0.36 & 0.45 & 0.04 & 0.01 & +0.41$^\dagger$ & +0.32$^\dagger$ \\
    Qwen3.8-Max & 0.90 & 0.93 & 0.55 & 0.21 & +0.38$^\dagger$ & +0.33$^\dagger$ & +0.68$^\dagger$ & 0.30 & 0.44 & 0.02 & 0.00 & +0.42$^\dagger$ & +0.27$^\dagger$ \\
    GLM-5.3 & 0.78 & 0.83 & 0.55 & 0.19 & +0.28$^\dagger$ & +0.37$^\dagger$ & +0.59$^\dagger$ & 0.11 & 0.20 & 0.04 & 0.01 & +0.17$^\dagger$ & +0.07$^\dagger$ \\
    GPT-6.1-Sol & 0.94 & 0.98 & 0.71 & 0.62 & +0.28$^\dagger$ & +0.09 & +0.32$^\dagger$ & 0.12 & 0.37 & 0.05 & 0.04 & +0.32$^\dagger$ & +0.07 \\
    \midrule
    Qwen2.5-7B & 0.22 & 0.23 & 0.07 & 0.02 & +0.16 & +0.05 & +0.20$^\dagger$ & 0.01 & 0.02 & 0.00 & 0.00 & +0.02 & +0.01 \\
    Llama-3.1-8B & 0.31 & 0.21 & 0.02 & 0.00 & +0.19$^\dagger$ & +0.02 & +0.30$^\dagger$ & 0.04 & 0.04 & 0.01 & 0.01 & +0.04 & +0.03$^\dagger$ \\
    Qwen2.5-14B & 0.44 & 0.42 & 0.19 & 0.04 & +0.23$^\dagger$ & +0.15 & +0.40$^\dagger$ & 0.06 & 0.04 & 0.01 & 0.01 & +0.03 & +0.05$^\dagger$ \\
    Qwen2.5-32B & 0.55 & 0.48 & 0.31 & 0.03 & +0.17 & +0.28$^\dagger$ & +0.52$^\dagger$ & 0.03 & 0.03 & 0.00 & 0.00 & +0.03 & +0.03 \\
    \bottomrule
  \end{tabular*}
\end{table*}

\subsection{Hospital: Damaged Identity Values}

Replacing Hospital's entity names with syllables reduces DeepSeek-V4-Pro's identity-column
repair F1 from 0.88 to 0.16. Qwen3.8-Max and GLM-5.3 have similarly large losses, giving a
0.59--0.72 range for these three models. Their W\,$-$\,S contrasts
are $+$0.33 to $+$0.37, in addition to positive O\,$-$\,W contrasts. F\,$-$\,W is also large
($+$0.28 to $+$0.45), with higher scores on fresh real records than on word-based names.
GPT-6.1-Sol's higher S score follows the construction regularities examined in
Section~\ref{sec:commercial}.

Detection remains stronger than repair (Table~\ref{tab:detection}). For example, DeepSeek
detects S identity errors with F1 0.73 but repairs them with 0.16, while its non-identity repair
scores remain between 0.91 and 0.95 across conditions. Name candidates can help resolve typos;
recovering an arbitrary identifier requires its mapping.
Open models show smaller O\,$-$\,S differences, from $+$0.20 to $+$0.52 at seed~0.

\begin{table}[t]
  \caption{Detection and repair diverge on surrogate identity columns. Entries give detection F1 /
  repair F1; detection counts corrupted cells found regardless of the value proposed for that cell.}
  \label{tab:detection}
  \small
  \setlength{\tabcolsep}{3pt}
  \begin{tabular*}{\columnwidth}{@{\extracolsep{\fill}}l ccc ccc@{}}
    \toprule
    & \multicolumn{3}{c}{Hospital} & \multicolumn{3}{c}{Movies} \\
    \cmidrule(lr){2-4}\cmidrule(l){5-7}
    Model & O & W & S & O & W & S \\
    \midrule
    DeepSeek & 1.0/.88 & .95/.50 & .73/.16 & .56/.36 & .41/.04 & .47/.01 \\
    Qwen3.8 & .98/.90 & .97/.55 & .72/.21 & .56/.30 & .57/.02 & .52/.00 \\
    GLM-5.3 & .98/.78 & .97/.55 & .72/.19 & .26/.11 & .12/.04 & .08/.01 \\
    GPT-6.1 & .95/.94 & .71/.71 & .65/.62 & .17/.12 & .11/.05 & .09/.04 \\
    \bottomrule
  \end{tabular*}
\end{table}

\finding{Syllable replacement lowers identity-column repair F1 by 0.59--0.72 for
DeepSeek-V4-Pro, Qwen3.8-Max and GLM-5.3 on Hospital. Detection remains stronger than repair.}\label{find:hospital}

\subsection{Movies: Candidates and Mappings}

\begin{figure}[t]
  \centering
  \includegraphics[width=\columnwidth]{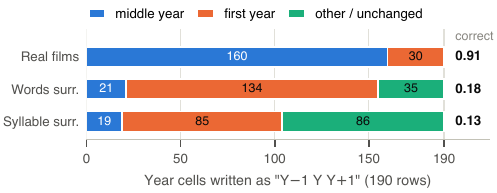}
  \caption{DeepSeek-V4-Pro's year repairs deteriorate after renaming, although all 190 year
  ranges are unchanged. Bars show the candidate written; the right panel shows repair accuracy.
  The middle year matches 167 references (88\%).}
  \label{fig:year}
  \Description{Stacked bars compare the year candidates selected on real and renamed films; accuracy falls after renaming although the candidate ranges are unchanged.}
\end{figure}

\textbf{Year ranges retain useful candidates.} In the 190 source-style rows, the year is
``Y$-$1 Y Y$+$1'' and the middle candidate matches 167 references (88\%). DeepSeek's year
accuracy is 91\% on real films, but only 18\% on W and 13\% on S, where it mostly chooses the first year
(Figure~\ref{fig:year}). Renaming shifts its choices from the middle toward the first candidate;
the prompt does not specify which candidate to select. \companionref{7} separates the 167
middle-reference cases from the other 23.
The other commercial models show the same pattern: 171--176 correct years on O, versus
24--48 on surrogates. Duration and genre repairs also decline after renaming
(Table~\ref{tab:moviescols}).

Rating-count formatting moves in the opposite direction. Its dirty and reference values are
identical across O/W/S, yet DeepSeek repairs 2 of 768 counts on O and 90 on S by removing
commas. All 90 S repairs occur in seven chunks. The rule remains locally available, but its
use varies across chunks and naming conditions.

\textbf{Identifiers require entity mappings.} All commercial models score at most 0.05 on
Movies' surrogate identity columns. Their
F\,$-$\,W gaps are $+$0.17 to $+$0.42, while the paired O\,$-$\,W contrasts are smaller
(Table~\ref{tab:main}). DeepSeek replaces 78 of 190 real slugs with the correct IMDb identifier
and none on surrogates (Table~\ref{tab:moviescols}). A fictitious slug does not determine its
random identifier, so the repairs on real films use mappings absent after renaming.

Without identifiers, F\,$-$\,W remains $+$0.26 to $+$0.33 across the four commercial models,
whereas O\,$-$\,W is $+$0.04 to $+$0.15. Original people-list errors often contain legitimate
Rotten Tomatoes credits, making the reference's source choice difficult even when the film is
known. GPT-6.1-Sol rarely guesses identifiers on O (10 of 190); its O\,$-$\,W of $+$0.07 is
not significant after correction, whereas its F\,$-$\,W gap is.

Open models repair at most 14 of 379 identity errors on O and three on W
(Table~\ref{tab:counts}). These near-zero results provide little evidence about their
response to entity renaming on Movies.

GLM's O\,$-$\,W remains significant after excluding incomplete replies:
$+$0.068 versus $+$0.070 on all chunks.

\begin{table}[t]
  \caption{Repaired cells and chunk wins:losses, excluding ties. Hospital compares W/S on 219 identity errors; Movies compares O/W on 379. Hospital chunk counts are descriptive; Movies sign tests use Holm correction over eight models.}
  \label{tab:counts}
  \small
  \setlength{\tabcolsep}{3.5pt}
  \begin{tabular*}{\columnwidth}{@{\extracolsep{\fill}}lrr rrr@{}}
    \toprule
    & \multicolumn{2}{c}{Hospital, W vs S} & \multicolumn{3}{c}{Movies, O vs W} \\
    \cmidrule(lr){2-3}\cmidrule(l){4-6}
    Model & W/S & Chunks & O/W & Chunks & $p$ \\
    \midrule
    DeepSeek & 104/28 & 35:1 & 97/11 & 35:3 & $<$.001 \\
    Qwen3.8 & 116/37 & 39:2 & 94/7 & 37:2 & $<$.001 \\
    GLM-5.3 & 117/32 & 39:2 & 25/8 & 16:3 & .022 \\
    GPT-6.1 & 120/100 & 19:6 & 24/10 & 13:6 & .334 \\
    Qwen-7B & 11/4 & 8:4 & 1/0 & 1:0 & 1 \\
    Llama-8B & 5/1 & 4:0 & 12/3 & 9:1 & .086 \\
    Qwen-14B & 35/7 & 23:2 & 14/3 & 10:0 & .012 \\
    Qwen-32B & 63/7 & 35:0 & 7/1 & 6:0 & .094 \\
    \bottomrule
  \end{tabular*}
\end{table}

\begin{table}[t]
  \caption{DeepSeek-V4-Pro Movies repairs by column (repaired / errors). Identity columns
  are listed first; each condition uses its own error counts.}
  \label{tab:moviescols}
  \small
  \setlength{\tabcolsep}{4pt}
  \begin{tabular*}{\columnwidth}{@{\extracolsep{\fill}}lrrrr@{}}
    \toprule
    Column & O & F & W & S \\
    \midrule
    identifier & 78 / 190 & 75 / 190 & 0 / 190 & 0 / 190 \\
    title & 0 / 17 & 4 / 17 & 1 / 17 & 0 / 17 \\
    director & 18 / 43 & 27 / 39 & 7 / 43 & 2 / 43 \\
    writers & 1 / 129 & 27 / 125 & 3 / 129 & 1 / 129 \\
    \midrule
    year & 173 / 190 & 183 / 190 & 35 / 190 & 25 / 190 \\
    duration & 90 / 187 & 101 / 187 & 42 / 187 & 41 / 187 \\
    genre & 44 / 180 & 70 / 167 & 3 / 180 & 1 / 180 \\
    rating & 30 / 261 & 73 / 280 & 9 / 261 & 9 / 261 \\
    rating count & 2 / 768 & 26 / 1000 & 24 / 768 & 90 / 768 \\
    \bottomrule
  \end{tabular*}
\end{table}

\finding{Movies' commercial-model F\,$-$\,W gaps are $+$0.17--$+$0.42 at seed~0, with
surrogate identity repair F1 at most 0.05. Year repair also deteriorates despite retained
candidate evidence.}\label{find:movies}

\subsection{Beers: Formatting Errors}

On Beers, we compare city/state repairs for real and fictitious names
(Table~\ref{tab:beers}). At seed~0, seven models show no significant positive O\,$-$\,S effect; their
contrasts range from $-$0.08 to $+$0.04, including GPT's $-$0.03 in the opposite direction.
Qwen3.8-Max instead repairs 122 of 127 cities on O, 108 on W and 82 on S, giving a significant
O\,$-$\,S difference of $+$0.24 (adjusted $p=0.042$). W repairs more cities than S in 15 chunks
and fewer in four. F is descriptive because its brewery concentration and alcohol distribution
differ (Section~\ref{sec:build-beers}).

\finding{Qwen3.8-Max has a positive O\,$-$\,S repair-F1 difference of $+$0.24 on Beers' formatting
errors; no significant positive effect appears in the other seven models.}\label{find:beers}

\begin{table}[t]
  \caption{Beers identity-column repair F1, O$-$S contrasts and repaired cities (O/F/W/S, of 127). $^\dagger$: Holm-corrected $p<0.05$ across eight O$-$S contrasts using entity/reply weights. F is descriptive; its brewery and alcohol-content distributions differ from O.}
  \label{tab:beers}
  \small
  \setlength{\tabcolsep}{3.5pt}
  \begin{tabular*}{\columnwidth}{@{\extracolsep{\fill}}lcccccr@{}}
    \toprule
    Model & O & F & W & S & O$-$S & Cities repaired \\
    \midrule
    DeepSeek-V4-Pro & 0.27 & 0.24 & 0.11 & 0.28 & $-$0.01 & 20 / 18 / 8 / 22 \\
    Qwen3.8-Max & 0.94 & 0.61 & 0.82 & 0.69 & +0.24$^\dagger$ & 122 / 67 / 108 / 82 \\
    GLM-5.3 & 0.94 & 0.88 & 0.89 & 0.94 & +0.00 & 120 / 110 / 116 / 117 \\
    GPT-6.1-Sol & 0.96 & 0.89 & 0.98 & 1.00 & $-$0.03 & 127 / 115 / 127 / 127 \\
    Qwen2.5-7B & 0.00 & 0.10 & 0.04 & 0.05 & $-$0.05 & 0 / 7 / 4 / 5 \\
    Llama-3.1-8B & 0.18 & 0.27 & 0.11 & 0.14 & +0.04 & 23 / 28 / 14 / 17 \\
    Qwen2.5-14B & 0.16 & 0.05 & 0.11 & 0.19 & $-$0.03 & 13 / 4 / 15 / 20 \\
    Qwen2.5-32B & 0.03 & 0.00 & 0.04 & 0.11 & $-$0.08 & 2 / 0 / 3 / 10 \\
    \bottomrule
  \end{tabular*}
\end{table}

\subsection{Stability over Construction Seeds}

Table~\ref{tab:seeds} reports construction variation separately from cluster intervals.
Hospital's O\,$-$\,S contrast varies by at most 0.06 per model, and W\,$-$\,S by at most
0.05. F\,$-$\,W is more sensitive to the fresh error draw, with ranges reaching 0.11 for
DeepSeek and 0.13 for open models. Movies' commercial-model gaps remain positive across seeds,
including $+$0.39 to $+$0.44 for DeepSeek.

\begin{table}[t]
  \caption{Identity-column contrast ranges over three construction seeds; seed 0 is in
  Table~\ref{tab:main}. Fresh clean records are fixed while errors are redrawn.}
  \label{tab:seeds}
  \small
  \setlength{\tabcolsep}{2.5pt}
  \begin{tabular*}{\columnwidth}{@{\extracolsep{\fill}}l ccc cc@{}}
    \toprule
    & \multicolumn{3}{c}{Hospital} & \multicolumn{2}{c}{Movies} \\
    \cmidrule(lr){2-4}\cmidrule(l){5-6}
    Model & F$-$W & W$-$S & O$-$S & F$-$W & O$-$W \\
    \midrule
    DeepSeek & .34--.45 & .33--.36 & .66--.72 & .39--.44 & .31--.32 \\
    Qwen3.8 & .34--.38 & .33--.38 & .68--.69 & .38--.42 & .26--.28 \\
    GLM-5.3 & .28--.29 & .34--.37 & .57--.61 & .14--.19 & .07--.08 \\
    GPT-6.1 & .28--.29 & .09--.10 & .32--.34 & .32--.35 & .07 \\
    \midrule
    Qwen-7B & .16--.22 & .02--.07 & .19--.20 & .01--.03 & .00--.01 \\
    Llama-8B & .14--.24 & .01--.04 & .30 & .03--.04 & .03 \\
    Qwen-14B & .23--.31 & .12--.17 & .37--.42 & .03--.06 & .05 \\
    Qwen-32B & .17--.30 & .27--.28 & .49--.52 & .02--.03 & .01--.03 \\
    \bottomrule
  \end{tabular*}
\end{table}

\section{Information Available for Repair}\label{sec:generalize}

\begin{figure*}[t]
  \centering
  \includegraphics[width=\textwidth]{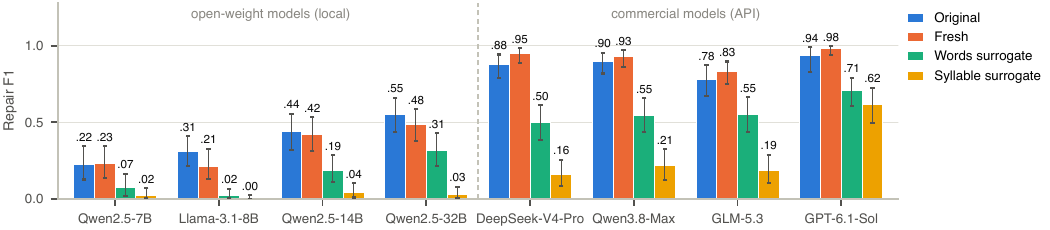}
  \caption{Hospital identity-column repair F1 across model sizes and families.
  Within Qwen2.5-7B/14B/32B, the O--S gap grows while S scores remain low.
  Llama and commercial models are shown for comparison; whiskers give 95\% cluster intervals.}
  \label{fig:scale}
  \Description{Condition scores across models show a larger original-to-syllable gap for larger Qwen models on Hospital while syllable-condition scores stay low.}
\end{figure*}

\subsection{Aligned Information and Repeated Records}\label{sec:labels}

\begin{figure}[tbp]
  \centering
  \includegraphics[width=\columnwidth]{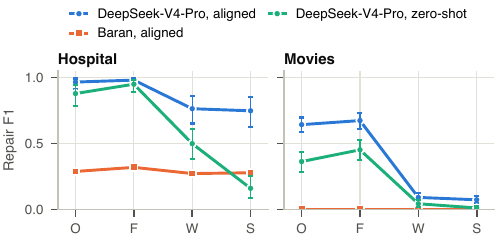}
  \caption{Identity-column repair F1 in zero-shot and aligned settings.
  DeepSeek shows 95\% cluster intervals; aligned Baran shows three-run means
  without error bars.}
  \label{fig:aligned}
  \Description{Hospital and Movies identity-column scores compare zero-shot and aligned DeepSeek with entity/reply cluster intervals. Baran points are means of three aligned runs, without intervals.}
\end{figure}

The aligned setting adds labeled target-value examples and error positions without enlarging
the main chunks. DeepSeek-V4-Pro improves across conditions but retains
F\,$-$\,W gaps of $+$0.22 on Hospital and $+$0.58 on Movies (Figure~\ref{fig:aligned}).
Hospital's rounded W and S scores differ by only 0.01, compared with the larger W\,$-$\,S
difference in the zero-shot setting.

Baran with the same chunks, labels and positions scores 0.27--0.32 on Hospital and zero on
Movies identity columns. With whole-table perfect detection, it reaches 0.91--0.93 and
0.25--0.27, respectively, with little condition variation (Table~\ref{tab:aligned}). Whole-table access exposes
repeated records that can supply values missing from the main chunks. DeepSeek still leads Baran in every aligned Hospital and
Movies condition, but its advantage narrows on surrogates.

\begin{table}[t]
  \caption{Identity-column repair F1 under aligned and whole-table information. Aligned uses the same chunks,
  20 labeled tuples and error positions; Baran is the mean of three runs. Whole-table Baran
  uses perfect detection. TabClean uses DeepSeek-V4-Pro with development rows excluded.}
  \label{tab:aligned}
  \small
  \begin{tabular*}{\columnwidth}{@{\extracolsep{\fill}}l cccc@{}}
    \toprule
    Method & O & F & W & S \\
    \midrule
    \multicolumn{5}{@{}l}{\emph{Hospital}} \\
    \quad DeepSeek, aligned & 0.97 & 0.98 & 0.76 & 0.75 \\
    \quad Baran, aligned & 0.29 & 0.32 & 0.27 & 0.28 \\
    \quad Baran, whole table & 0.91 & 0.93 & 0.93 & 0.93 \\
    \quad TabClean & 1.00 & 1.00 & 0.96 & 0.99 \\
    \multicolumn{5}{@{}l}{\emph{Movies}} \\
    \quad DeepSeek, aligned & 0.64 & 0.67 & 0.09 & 0.07 \\
    \quad Baran, aligned & 0.00 & 0.00 & 0.00 & 0.00 \\
    \quad Baran, whole table & 0.27 & 0.26 & 0.26 & 0.25 \\
    \quad TabClean & 0.00 & 0.00 & 0.03 & 0.00 \\
    \multicolumn{5}{@{}l}{\emph{Beers}} \\
    \quad DeepSeek, aligned & 1.00 & 1.00 & 1.00 & 1.00 \\
    \quad Baran, aligned & 0.19 & 0.77 & 0.20 & 0.22 \\
    \quad Baran, whole table & 0.94 & 1.00 & 0.94 & 0.96 \\
    \quad TabClean & 1.00 & 1.00 & 1.00 & 1.00 \\
    \bottomrule
  \end{tabular*}
\end{table}

\subsection{Model Scale}\label{sec:scale}

Within Qwen2.5, Hospital's O\,$-$\,S grows from $+$0.20 at 7B to $+$0.40 at 14B and $+$0.52
at 32B, in the same order at every construction seed (Figure~\ref{fig:scale};
Table~\ref{tab:seeds}). S repair F1 stays low at 0.02, 0.04 and 0.03; the increase is mainly
in scores on the original hospital names.

\subsection{Recoverable Syllable Names}\label{sec:commercial}

GPT-6.1-Sol repairs 100 of Hospital S's 219 identity errors with 105 edits, compared with
28--37 repairs for the other commercial models. The constructor's 24-syllable inventory is
represented in every chunk and permits a
unique restoration for 95 corrupted names. GPT restores 74 of them; the other commercial models
restore 8--12 and open models at most one (Table~\ref{tab:decoding}). The inventory is used by
our analysis, not supplied to the cleaner. On digits, which it does not constrain, GPT repairs
6 of 97. Its S advantage is concentrated on names recoverable from the restricted vocabulary.

\begin{table}[t]
  \caption{Hospital S repairs by recovery category. Unique and Ambig. use the 24-syllable inventory;
  Generic covers English terms, and Digits covers numeric tokens. Headers give cell counts, covering
  217 of 219 errors, with two address-spacing errors excluded from these counts.}
  \label{tab:decoding}
  \small
  \setlength{\tabcolsep}{4pt}
  \begin{tabular}{@{}lrrrr@{}}
    \toprule
    Model & Unique (95) & Ambig. (11) & Generic (14) & Digits (97) \\
    \midrule
    DeepSeek-V4-Pro & 8 & 0 & 13 & 5 \\
    Qwen3.8-Max & 12 & 0 & 12 & 11 \\
    GLM-5.3 & 10 & 0 & 12 & 8 \\
    GPT-6.1-Sol & 74 & 4 & 14 & 6 \\
    Open models & 0--1 & 0 & 0--3 & 0--6 \\
    \bottomrule
  \end{tabular}
\end{table}

\subsection{Program-Generating Cleaner}\label{sec:e4}

TabClean~\cite{tabclean2026} generates cleaning programs. Its identity-column scores stay near
perfect on Hospital and Beers, but near zero on Movies (Table~\ref{tab:aligned}). For example,
its Beers output splits ``San Francisco CA'' into city and state, both retained in the input.
Movies' source mappings and omitted credits require external evidence.

\finding{Labels and error positions improve DeepSeek-V4-Pro's repair without closing its
F\,$-$\,W gaps. Whole-table Baran and TabClean remain strong under renaming on Hospital
and Beers, but weak on Movies.}\label{find:general}

\section{Preserving Assigned Values}\label{sec:conflict}

Movies often pairs legitimate values from different sources, so a familiar credit alone does
not identify the task's target. We next fix a value to preserve and test whether a familiar
name draws the cleaner toward the real-world fact.

\subsection{Constructing Knowledge Conflicts}\label{sec:conflictdesign}

Starting from S, we assign target cells another real director or location in both dirty and
reference tables. Correctness here means preserving the assigned target, although other cells
in the row may still contain errors. Table~\ref{tab:conflictexample} illustrates the assignment
of Stanley Kubrick to \emph{Taken}, whose real director is Pierre Morel.

\begin{table}[t]
  \caption{A conflict target with the same assigned Director reference in three name versions.
  The real credit of \emph{Taken} (2008) is Pierre Morel; all other cells remain identical.}
  \label{tab:conflictexample}
  \small
  \begin{tabularx}{\columnwidth}{@{}lXll@{}}
    \toprule
    Version & Name & Year & Director \\
    \midrule
    Exact name & Taken & 2008 & Stanley Kubrick \\
    One-letter name & Takun & 2008 & Stanley Kubrick \\
    Surrogate control & Drikel Joujou & 2008 & Stanley Kubrick \\
    \bottomrule
  \end{tabularx}
\end{table}

Each target has an exact real name, a one-vowel-changed name and its syllable control. Only
the name differs across versions; attributes, other values, error positions and chunks are
fixed. Hospital and Beers receive internally consistent locations. Movies also updates
related director occurrences consistently, preventing another column from supplying the old
director as a candidate. \companionref{2} specifies these cross-column replacements.

The common pool contains 57 films, 24 hospitals and 48 breweries known to DeepSeek-V4-Pro.
Main chunks contain three targets and 17 other-entity fillers, with one row per target.
Supplementary Hospital/Beers conditions use three rows per target. Every model receives the
same pool, but rates use its own subset passing separate factual queries: 52--57 films,
22--24 hospitals and 7--48 breweries. The factual screen for films accepts a credited full name
or a surname of at least four letters; the location screen requires matching city and state.
Rates are conditional on the fixed pool.

The standard prompt asks for cleaning while preserving correct cells, without specifying
source priority. The \emph{table-evidence prompt} explicitly favors input-table evidence over
real-world knowledge. Both use the same chunks and known subsets.

\subsection{Measuring Rewrites}\label{sec:conflictmetric}

A directed rewrite changes the assigned Director or City to the paired entity's real-world
fact. Director outputs must contain a full credited name; cities must match the normalized
real city. The one-letter-name variant instead measures restoration of the exact real name.
These events are scored separately from any change to the target cell. Output parsing and
matching normalization are specified in \companionref{2}.

Let $E_m$ contain entities passing model $m$'s factual queries, $s$ be target rows per entity,
and $z_{ej}$ indicate a matching rewrite in row $j$ of entity $e$. We report
\begin{equation*}
  \widehat{r}_m = \frac{1}{|E_m|}\sum_{e\in E_m}
                 \frac{1}{s}\sum_{j=1}^{s} z_{ej}.
\end{equation*}
For Movies, $s=1$, giving rewritten known films divided by known films. For $s=3$, row indicators
are first averaged within entities. Targets without a matching correction, including unusable
replies, remain in the denominator. Surrogate controls use the same known subset and paired
fact. Reply-cluster intervals follow Section~\ref{sec:setup}.

\begin{figure}[t]
  \centering
  \includegraphics[width=\columnwidth]{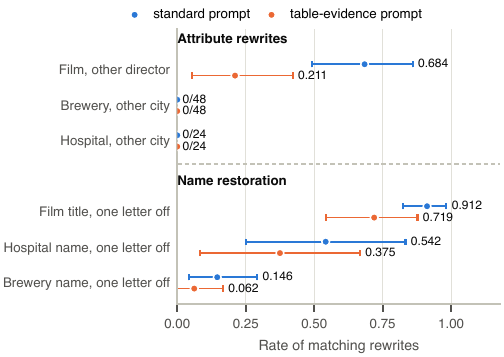}
  \caption{DeepSeek-V4-Pro rewrites assigned directors more often than cities. Attribute rewrites
  and name restoration are separate events, using 57 films, 24 hospitals or 48 breweries with
  one row per target. Whiskers give 95\% reply-cluster intervals; all-or-none outcomes show counts.
  Controls have zero directed rewrites.}
  \label{fig:near}
  \Description{Paired points compare standard and table-evidence prompts for director or city rewrites and for restoration of one-letter-changed entity names.}
\end{figure}

\begin{figure}[t]
  \centering
  \includegraphics[width=\columnwidth]{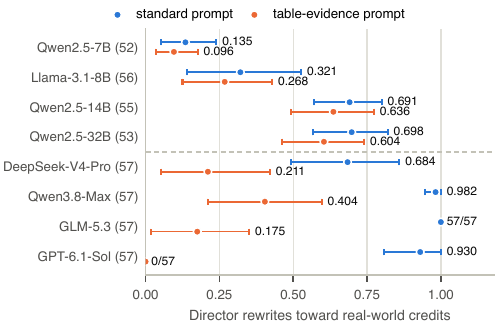}
  \caption{Director rewrites under standard and table-evidence prompts. Parentheses give
  known-film counts shared by both prompts; Table~\ref{tab:conflictcounts} includes control edits.
  Whiskers give 95\% reply-cluster intervals, with counts for all-or-none outcomes.}
  \label{fig:nearmodels}
  \Description{Per-model director rewrite rates and intervals compare the two prompts; labels give each model's known-film denominator.}
\end{figure}

\subsection{Results}\label{sec:conflictresults}

\textbf{Explicit table priority.} With the table-evidence prompt, GPT does not replace any
assigned director with a real-world credit (0/57). DeepSeek still rewrites 12 of 57 directors; GLM and
Qwen3.8-Max rewrite 10/57 and 23/57, respectively. Rates for Qwen2.5 at 14B and 32B remain
above 0.60 (Table~\ref{tab:conflictcounts}). These updates contradict the prompt's explicit
instruction to use table evidence.

\textbf{Director rewrites.} Commercial models know the same 57 films and
rewrite 68\%--100\% of their assigned directors under the standard prompt
(Figure~\ref{fig:nearmodels}). DeepSeek, for example, rewrites 39/57. Controls also receive
Director edits, but none matches the paired film's real credit (Table~\ref{tab:conflictcounts}).

\begin{table}[t]
  \caption{Director rewrite counts among known films ($n/N$). Exact-name targets count edits
  matching a real-world director credit. Surrogate controls count any Director change;
  none matches the paired film's real credit.}
  \label{tab:conflictcounts}
  \small
  \setlength{\tabcolsep}{3pt}
  \begin{tabular*}{\columnwidth}{@{\extracolsep{\fill}}lrrrr@{}}
    \toprule
    & \multicolumn{2}{c}{Exact-name target} & \multicolumn{2}{c}{Surrogate control} \\
    \cmidrule(lr){2-3}\cmidrule(lr){4-5}
    Model & Standard & Evidence & Standard & Evidence \\
    \midrule
    DeepSeek-V4-Pro & 39/57 & 12/57 & 2/57 & 0/57 \\
    Qwen3.8-Max & 56/57 & 23/57 & 14/57 & 3/57 \\
    GLM-5.3 & 57/57 & 10/57 & 10/57 & 1/57 \\
    GPT-6.1-Sol & 53/57 & 0/57 & 7/57 & 0/57 \\
    Qwen2.5-7B & 7/52 & 5/52 & 0/52 & 0/52 \\
    Llama-3.1-8B & 18/56 & 15/56 & 2/56 & 2/56 \\
    Qwen2.5-14B & 38/55 & 35/55 & 2/55 & 0/55 \\
    Qwen2.5-32B & 37/53 & 32/53 & 0/53 & 0/53 \\
    \bottomrule
  \end{tabular*}
\end{table}

Within Qwen2.5, conditional rates rise from 0.135 at 7B to 0.691 at 14B and 0.698 at 32B.
Their known subsets differ, as shown in Table~\ref{tab:conflictcounts}. Llama knows 56 films
and rewrites 18 (0.321).

\textbf{Director versus city.} Standard-prompt city rewrites reach at most
0.042 on Hospital and 0.140 on Beers across models.
DeepSeek also restores 91\% of one-letter-changed film names (Figure~\ref{fig:near}). This may
correct spelling without changing the assigned attribute.

\finding{Under explicit table-evidence instructions, commercial models' director rewrite
rates range from 0.000 to 0.404; Qwen2.5-14B/32B remain above 0.60. Standard-prompt rates for
commercial models are 68\%--100\%, while city rewrites are less frequent in these tests.}\label{find:conflict}

\section{Benchmark Quality}\label{sec:quality}

Rayyan contains systematic date and page conversions in its reference values. A cleaner can
receive credit for reproducing these conversions but none for an externally matched
ISSN correction.

\subsection{Systematic Reference Transformations}

Rayyan's audit pair has 1,000 rows and eleven columns, differing in 948 cells. Its
values exhibit patterns characteristic of spreadsheet conversion~\cite{ziemann2016gene}.
All 722 date differences map input \texttt{m/d/yy} to \texttt{d/int(yy)/mm}, with a two-digit
input month in the final position: for example, \texttt{1/1/13} becomes \texttt{1/13/01}.
Similarly, 31 of 32 pagination differences turn page ranges into month-name strings, such as
\texttt{70-6} to \texttt{Jun-70}.

\subsection{What the Reference Rewards}\label{sec:rayyan-rewards}

On the audit input, DeepSeek-V4-Pro's zero-shot output scores F1 0.002. It changes dirty ISSN
\texttt{Mar-22} to \texttt{0003-3022}, matching a title-matched PubMed record, but the reference
is \texttt{22-Mar}. This correction receives no repair credit.

TabClean's distributed Rayyan input has 960 difference cells, rather than the audit's 948.
Both are 1,000-row, eleven-column tables with the same date-conversion pattern. We score the
archived output against its actual input pair and exclude 150 development rows. The remaining
850 rows contain 808 errors; 498 effective edits yield 461 credited repairs, giving
$F_1=2\times461/(498+808)=0.71$ after rounding. Of these credited repairs, 434 (94\%) reproduce
the reference date transformation. \companionref{3} specifies the actual inputs, development
indices and output.

\subsection{Stratified PubMed Checks}\label{sec:rayyan-audit}

We sample up to 80 cells per error type, including all volume/issue, pagination and ISSN
differences, for 279 cells. Automatic title matching selects the highest-similarity PubMed
candidate among up to five results at a normalized similarity of at least 0.9. This yields
152 matched cells associated with 121 PMIDs: 54.5\% overall and 47.5\%--61.3\% across types
(Table~\ref{tab:rayyan}). Query construction and retrieval provenance appear in \companionref{3}.

\textbf{Field-specific checks.} The date field \texttt{article\_jcreated\_at} is checked for
agreement with PubMed's publication year; the dataset does not document its exact semantics.
Volume and issue checks measure presence, using reference \texttt{-1} and dirty empty strings
as absence. Two reference volumes of \texttt{1}, for example, pass presence while differing
numerically from saved PubMed values \texttt{31} and \texttt{21}. Pagination matches the page
string, and ISSN accepts print or electronic identifiers. For encoding differences, a nonempty
set of reference-side changed words must all occur in the case-folded PubMed
word set, with accents retained. Encoding's dirty side is unassessed.

\begin{table}[tbp]
  \caption{Rayyan field checks: population $N_h$, sample $n_h$, title matches $m_h$,
  failed reference and passed dirty checks per matched cell. A dash denotes an unassessed dirty check.}
  \label{tab:rayyan}
  \small
  \setlength{\tabcolsep}{3pt}
  \begin{tabular*}{\columnwidth}{@{\extracolsep{\fill}}lrrrcc@{}}
    \toprule
    Check & $N_h$ & $n_h$ & $m_h$ & \shortstack{Reference fail\\$b_h/m_h$} & \shortstack{Dirty pass\\$d_h/m_h$} \\
    \midrule
    Year & 722 & 80 & 44 & 43/44 & 42/44 \\
    Changed words & 107 & 80 & 38 & 38/38 & --- \\
    Vol./issue & 75 & 75 & 46 & 28/46 & 17/46 \\
    Pages & 32 & 32 & 18 & 18/18 & 17/18 \\
    ISSN & 12 & 12 & 6 & 6/6 & 0/6 \\
    \bottomrule
  \end{tabular*}
\end{table}

Let $N_h$ be the error-cell population of type $h$, $m_h$ its title-matched count, and $b_h$
the matched references failing that field's check. The weighted reference-check failure rate is
\begin{equation}
  \widehat q = \sum_h \frac{N_h}{948}\frac{b_h}{m_h}.
  \label{eq:rayyan-weighted}
\end{equation}
Weighting the five field checks by the 948-cell audit composition gives a reference-check
failure rate of 95.2\% (95\% interval: 91.2--97.8\%; Table~\ref{tab:rayyan}). Each within-type
proportion uses title-matched cells. Dates account for 722 of 948 error cells and most of the
weight. The interval conditions on the saved title matches and field checks;
matching uncertainty and dependence between cells sharing a PMID are outside this estimate.

Among the 44 title-matched date cells, 42 dirty values agree with PubMed's publication year,
whereas 43 reference values do not. These sampled publication-year checks differ from
TabClean's credited-conversion count of 434/461.

\subsection{Reference Issues in Other Benchmarks}

Hospital's reference places Helen Keller Memorial Hospital in Jefferson County, although its
city, Sheffield, is in Colbert County. Three commercial models correct some of these cells and
are penalized as false edits. Counts and duplicate writer credits in Movies are detailed in
\companionref{3}.

\finding{On Rayyan, 434/461 of TabClean's credited repairs (94\%), after excluding 150
development rows, reproduce the reference date transformation. An externally matched ISSN
correction earns no repair credit.}\label{find:quality}

\section{Lessons and Recommendations}\label{sec:discussion}

\textbf{Distinguish missing mappings from retained candidates.} An identifier may lose its mapping after
renaming, while a year range retains its candidates (Section~\ref{sec:recoverability}). Report
formatting repairs separately once the target convention is established: a local conversion
and a source-specific value pose different recovery problems
(Findings~\ref{find:hospital}--\ref{find:beers}). If a renamed identifier has no mapping, supply a
candidate list or retrieval source before comparing a method's ability to recover it. Where
candidates remain, report which candidates the method selects, as in the year comparison,
rather than interpreting all score losses as missing external knowledge.

\textbf{Align labels, error positions and supporting records.} Compare chunk-based and
whole-table methods under aligned labels and error positions (Finding~\ref{find:general}).
Report supporting-record access separately: another record of the same entity may supply a
replacement absent from the chunk. Equal detection information does not imply equal evidence
for recovering values.

\textbf{Set source priority and test value preservation.} Movies requires selecting a preferred
source, whereas the conflict test requires retaining an assigned director. Count factual rewrites
separately from other edits: a real credit can still be the wrong update for that task
(Finding~\ref{find:conflict}). For cross-source errors, name the preferred source explicitly:
a Rotten Tomatoes credit can be legitimate yet differ from an IMDb reference. For assigned
values, state that the table takes priority and include unchanged target cells in evaluation.

\textbf{Inspect the edits rewarded by the reference.} Rayyan credits date conversion but rejects an
externally matched ISSN. Compare fields with independent records under explicit criteria, and
report the scored input and development exclusions (Finding~\ref{find:quality}). Separate
mechanical conversions from changes in field content. An external year check also requires
the date field's semantics, while a weighted summary requires the match coverage and criteria
for each field. Inspecting high-credit edits is especially useful when a single transformation
accounts for most repaired cells.

\subsection{Applying \core{} to Other Benchmarks}\label{sec:apply}

Applying \core{} requires profiling errors and reference conventions, selecting a compatible
fresh source, and building the two naming styles with acceptance checks. Report F1 contrasts
alongside error-type counts and candidate selections, as in the Movies year
analysis. Record which field relationships survive identity replacement: geographic
codes can retain familiar meanings after names change. \companionref{4} describes the
adaptation procedure.

\subsection{Artifact and Result Replay}\label{sec:artifact}

The statistical core replays condition contrasts, conflict rates and Rayyan checks offline from
Hospital/Beers tables, Movies numerical witnesses and re-scored continuation counts.
Authorized local validation checks historical movie inputs, replies and scoring-code hashes
before re-scoring. The repository \texttt{README} gives the commands;
\companionref{5} maps each analysis to its inputs, scoring code and outputs;
\companionref{6} gives the extended displays.

\subsection{Limitations}\label{sec:limitations}

The datasets diagnose recovery mechanisms and reference transformations in entity-rich public
tables, rather than estimate their prevalence across cleaning workloads. Transfer to private data
also depends on size, sparsity, task complexity and the availability of supporting records.
Our comparisons change name forms as well as familiarity,
and F introduces record-difficulty differences (Table~\ref{tab:conditions}). Beers F has
different brewery concentration and alcohol distributions, while Hospital's restricted syllables
provide the regularity observed for GPT (Section~\ref{sec:commercial}). The study covers eight
commercial and open-weight cleaners on English tables concerning mainly US entities, with
decoding settings in \companionref{2}; it does not cover Claude or Gemini.

The intervals describe variation across entities and replies within fixed constructions.
New backgrounds, regrouped prompts and repeated API calls introduce other sources of variation.
\companionref{7} gives the entity--reply structure and shows how repair F1 changes when one
hospital's rows are omitted from the saved outputs.

\section{Conclusion}\label{sec:conclusion}

Renaming Movies lowers DeepSeek-V4-Pro's year accuracy from 91\% to 18\% even though the
candidate years remain in the input. Identifier repairs decline too, but their mappings are
no longer available. On Hospital, GPT repairs 74 of 95 errors with a unique restoration under
the constructor's restricted syllable inventory. The paired comparisons in \core{} retain
error positions, while fresh records
and a screened continuation probe examine transfer and benchmark-row memory separately.
The probe finds no positive evidence of row memory in its tested models.

The assigned-value and reference analyses reveal errors that conventional repair scores can
miss. Some models replace assigned directors with real credits even after an explicit
table-evidence instruction. On Rayyan, 434/461 of TabClean's credited repairs reproduce a
reference date transformation, while an externally matched ISSN correction earns no credit.
For evaluation, these cases call for tests with a specified source priority and inspection of
the edits counted as correct. The constructions and scoring code support comparisons with
other data cleaners.

\clearpage
\nobalance
\bibliographystyle{ACM-Reference-Format}
\bibliography{refs}

\end{document}